\newif\ifdraft
\drafttrue

\documentclass[11pt]{article}
\usepackage[utf8]{inputenc}
\usepackage{graphicx}
\usepackage[hscale=0.7,vscale=0.8]{geometry}
\usepackage{amsmath}
\usepackage{amssymb}
\usepackage{verbatim}
\usepackage{soul}
\usepackage{url}
\usepackage{xcolor}
\usepackage{textcomp}

\begin{document}

\title{The SFS Summer Research Study at UMBC:\\
Project-Based Learning Inspires Cybersecurity Students}


\author{Alan Sherman,$^1$
Enis Golaszewski,$^1$
Edward LaFemina,$^1$\\
Ethan Goldschen,$^1$
Mohammed Khan,$^3$
Lauren Mundy,$^2$
Mykah Rather,$^3$\\
Bryan Solis,$^2$
Wubnyonga Tete,$^3$
Edwin Valdez,$^2$
Brian Weber,$^1$\\
Damian Doyle,$^1$
Casey O'Brien,$^3$
Linda Oliva,$^1$
Joseph Roundy,$^2$
Jack Suess,$^1$\\
Cyber Defense Lab\\
University of Maryland, Baltimore County (UMBC)\\
Baltimore, Maryland 21250\\
email: \{sherman, golaszewski\}@umbc.edu
}

\date{{\today}}  

\maketitle

\footnotetext[1]{University of Maryland, Baltimore County}
\setcounter{footnote}{1}
\footnotetext[2]{Montgomery College}
\setcounter{footnote}{2}
\footnotetext[3]{Prince George's Community College}
\setcounter{footnote}{3}

\begin{abstract} 

May 30-June 2, 2017, 
Scholarship for Service (SFS) scholars at the University of Maryland, Baltimore County (UMBC) 
analyzed the security of a targeted aspect of the UMBC computer systems.  
During this hands-on study, with complete access to source code, students 
identified vulnerabilities, devised and implemented exploits, and suggested mitigations.  
As part of a pioneering program at UMBC to extend SFS scholarships to community colleges,
the study helped initiate six students from two nearby community colleges, who transferred to
UMBC in fall 2017 to complete their four-year degrees in computer science and information systems.

The study examined the security of a set of 
``NetAdmin'' custom scripts that enable UMBC faculty and staff to open the
UMBC firewall to allow external access to machines they control for research purposes.
Students discovered vulnerabilities stemming from 
weak architectural design, record overflow, and failure to sanitize inputs properly. 
For example, they implemented a record-overflow and code-injection exploit 
that exfiltrated the vital API key of the UMBC firewall.

This report summarizes student activities and findings, and reflects on lessons learned
for students, educators, and system administrators.
Our students found the collaborative experience inspirational;
students and educators appreciated the authentic case study; and 
IT administrators gained  access to future employees and 
received free recommendations for improving the security of their systems.
We hope that other universities can benefit from our motivational and educational strategy of
teaming educators and system administrators
to engage students in active project-based learning 
centering on focused questions about their university computer systems.

\end{abstract}

\bigskip
\noindent {\bf Keywords.}
Code injection,
computer and network security,
cybersecurity,
CyberCorps: Scholarship for Service (SFS),
firewalls,
NetAdmin,
project-based learning,
record overflow,
security evaluation, 
UMBC SFS Summer Research Study.

\clearpage
\section{Introduction} 
\label{sec:intro}

During four summer days in 2017, cybersecurity students at 
the University of Maryland, Baltimore County (UMBC)
analyzed the security of a targeted aspect of their university computer systems.
We report on this novel summer research study, its technical findings, and 
takeaways for students, educators, and system administrators.

In fall 2016, UMBC was one of ten
schools that pioneered a new strategy for attracting talented cybersecurity professionals
into government service: extend {\it CyberCorps: Scholarship for Service (SFS)} scholarships
to nearby partnering community colleges (CCs).  There are outstanding students at CCs, and
for many of them the financial challenges of attending a four-year college full-time are daunting.
With support from the National Science Foundation (NSF), UMBC offered six students the following 
contract:  after completing an associate's degree, transfer to UMBC and complete a bachelor's degree 
in a cybersecurity-related major.  In return for three years of generous support 
(tuition, fees, health insurance, stipend, and more) starting in the last year of CC, students work
for government (federal, state, local, or tribal) for each year of support.
UMBC inducted three students from Montgomery College (MC) and three from Prince George's Community College (PGCC),
who all now are pursuing degrees at UMBC in computer science, computer engineering, or information systems.

To integrate these students into the existing UMBC SFS cohort, to enhance their cybersecurity knowledge and skills,
and to inspire them, Alan Sherman organized a four-day SFS Summer Research Study at UMBC for all SFS scholars at
MC, PGCC, and UMBC.  Sherman also invited professors, researchers, 
and graduate students from UMBC, and personnel from the
National Security Agency (NSA), to interact with the students as technical experts.
Everyone worked collaboratively on the same focused practical problem: analyze the security
of the NetAdmin capability of the UMBC computer systems, which through a web interface enables UMBC faculty and staff to
open the UMBC firewall to allow external access to machines they control for research purposes.  
The project enjoyed strong cooperation from UMBC's Division of Information Technology (DoIT), 
which is responsible for managing UMBC computer systems in support of teaching, research, and administration.
Participants were given access to all relevant source code.  
The students explored all aspects of the problem, including adversarial models, 
architectures, key management, use of cryptography, 
authentication and key-establishment protocols, 
software implementations, configurations, and policy.

By the end of the first day, the SFS scholars identified several potential vulnerabilities and began devising
exploits, working safely in a virtual copy provided by DoIT of the actual network.  
Throughout and after the study, students presented their findings to DoIT and made several recommendations
for mitigating the issues they discovered. 
Particularly interesting and illuminating were the end-of-day discussions between students and DoIT staff
(including the primary NetAdmin script author) during which each side shared their perspectives on the situation,
how the vulnerabilities arose, the risks they pose, 
how to deal with them, and how to improve DoIT's processes.
  
Although the ideas underlying the attacks are not new, the analysis of UMBC's \hbox{NetAdmin} is.
More importantly, we hope that other students, educators, and system administrators 
may learn from our experiences in collaborative project-based learning.
All of our students reported that the summer study inspired them and enhanced their knowledge and skills.
Students and educators appreciated the authentic case study.
UMBC's system administrators gained access to highly qualified potential student employees (several of whom
now work for DoIT) and received free consulting on how to improve the security of their system.
There are many benefits for educators to partner with their university's IT department 
to use their university's computing systems as a cybersecurity learning laboratory.

UMBC is a midsize public university that emphasizes science and technology.  
Recognized as a National Center of Academic Excellence in Cyberdefense Education and Research (CAE, CAE-R), 
UMBC offers undergraduate and graduate tracks in cybersecurity leading to BS, MS, and PhD degrees
in computer science and information systems, 
and the MPS degree in cybersecurity.
In 2017, it won first place at the National Collegiate Cyberdefense Competition.

We assume the reader is familiar with the basics of cybersecurity, as introduced, for example,
by Kim and Soloman~\cite{kim14}, Bishop~\cite{bishop03}, and Sherman, et al.~\cite{exploring17}.

\section{The SFS Summer Study at UMBC} 
\label{sec:study}

This section identifies some of the key decisions we faced in organizing the
summer study and our rationale behind the choices we made.

A summer study appealed to us because it emphasized 
collaboration, problem solving, and independent thinking
in addressing an important, practical, challenging problem involving a variety of issues.
Cybersecurity is a broad discipline in which it is essential 
to form teams with appropriate diverse skills and talents.  
Our collaborative approach resonated with the cohort philosophy of the SFS program.
Inspiration also came from the SCAMP summer studies organized by
the Institute for Defense Analyses, in which a team collaboratively works on one or two 
research problems.
 
We reasoned that the most effective way to initiate new SFS scholars (especially from CCs)
into UMBC's SFS cohort
was to engage them in solving a technical research problem. 
We decided to remain focused on this task and not to spend any time
on anything else, such as informational presentations about UMBC.

We sought a problem that was rich and challenging, yet tractable in one week.
Choosing a problem that directly benefited the UMBC community was a major benefit,
promoting the service tenent of the SFS program.  
During the academic year, each SFS scholar at CC took a special
research course through which they solved IT security problems for their county government.
Analyzing the security of a focused aspect of the UMBC computer systems offered
all the elements we sought.

Focusing on custom scripts written at UMBC provided many attractive properties.
First, it was essential that the students be given all source code. 
There would be much greater
legal and administrative barriers to providing source code written by third parties.
Access to source code was important because security should not be based on
the obscurity of the NetAdmin process, and we assume that an adversary could obtain the source code.
Second, the NetAdmin scripts are short; 
by contrast, dealing with UMBC's huge commercial products (e.g., PeopleSoft) would be difficult. 
Third, these scripts had never undergone any critical technical security evaluation, 
and the likelihood of there being vulnerabilities seemed high.

Given the sensitive nature of problem, each participant signed a non-disclosure agreement with DoIT.   
Students acted responsibly and responded positively to the trust bestowed on them.
For DoIT, it was easy to trust the students because 
they were carefully vetted into the SFS program,
some held top-secret clearances, and
any malicious behavior would jeopardize their scholarships.

DoIT staff gave students all relevant system information---including source code, 
scripts, network architectures, protocols, and configurations---and 
were available to answer questions about them.
They also provided a virtual machine of the relevant parts of the campus network,
together with the Kali Linux suite of cybersecurity tools.

The diverse in-person participants comprised six CC transfer students, 
three UMBC undergraduates, and one PhD student.  The problem-solving process was entirely student driven.
Two UMBC professors and two NSA experts
visited each day, to answer any questions about technical subjects or
use of tools. 

The study took place from 9~a.m. to 5~p.m. Tuesday through Friday in a large room with tables, whiteboard, and projector. 
Basing our approach on the tenets of Project Based Learning (PBL)~\cite{blumenfeld91},
we presented the challenge and tasked the team
to create a structure that would achieve the desired goals of analyzing the security of NetAdmin. 
We encouraged students to approach the project in any way they saw fit.  
Replicating authentic scenarios of work places, 
our student-centered approach 
forced students to negotiate group dynamics and to apply their technical skills while fully engaging in a real-life task.  
This structure supported sustained inquiry and reflection.  

Students self-selected into teams, each exploring some focused aspect of the problem.
For example, one team surveilled the landscape, identifying the network topology,
operating system, machines, and versions of component software.
Other teams focused on architectural issues, source code analysis,
and known vulnerabilities of the operating system and component software in use.
More experienced students emerged as leaders.

Late each afternoon, representatives from DoIT, including the primary NetAdmin script author,
joined the group for a discussion of the day's findings.
To involve students who were unable to attend in person (e.g., due to required internships
some of which were out-of-state), one of the students led a one-hour evening chat session on Goggle Hangouts.

PBL is an instructional approach in which small groups of students 
engage in authentic tasks and learning occurs through their consideration of relevant problems. 
Students pursue solutions by asking and revising questions, debating ideas, generating predictions, 
experimenting, collecting data, drawing conclusions, communicating their ideas and findings, 
refining approaches, and creating products~\cite{blumenfeld91}.
 
PBL holds great promise in cybersecurity because there is a proliferation 
of complex problems in the field and projects can support students to sustain their effort and direct their learning. 
PBL supports students to develop diverse approaches to solving real-world problems.  
Students are task focused and they can try out a variety of solutions 
and receive timely feedback on their approaches. 
They engage in collaboration and reflection that deepen their learning and enhance the transferability of skills.

There are many examples of PBL in cybersecurity (e.g., NJIT's Cyber-RWC Summer
Camp\footnote{\url{https://sci.njit.edu/gencyber/RWCCybersecurityCampBrochure-Summer2016.pdf }}), 
though there are relatively few scholarly articles on this subject (e.g., Conklin and White~\cite{conk05},
whose graduate course includes some elements similar to our study).
We are strong believers in the value of PBL, 
as evidenced by our participation in the INSuRE Project~\cite{insure17}.

\section{The Problem} 
\label{sec:problem}

We define our research problem by describing
the UMBC campus network,
the origin and use of the NetAdmin tool,
our research questions, and
our adversarial model.
In short, our task is to analyze the security implications
of using the NetAdmin tool and to make appropriate recommendations to DoIT.

\begin{figure}
\centering
\includegraphics[scale=1, bb=0 0 100 100]{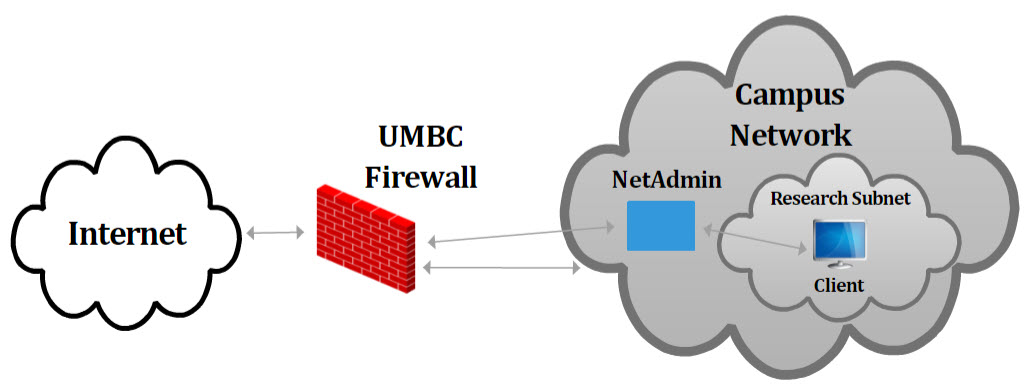}
\caption{Architecture of the UMBC network including the NetAdmin tool,
which is accessible to machines on the research subnet.}
\label{fig:arch}
\end{figure}

\subsection{The UMBC Campus Network}
\label{sec:network}

UMBC maintains a Palo Alto firewall on the network boundary that filters traffic
originating from, or destined for, addresses residing outside of the campus network.
All traffic between the Internet and the UMBC network passes through the UMBC firewall.
As shown in Figure~\ref{fig:arch}, the NetAdmin server is behind the UMBC firewall
and is accessible from machines located on the research subnet or via certain authenticated
VPN connections.  
The research subnet connects faculty-operated servers and machines.
All users authenticate to the system through a single sign-on myUMBC Sibboleth service.

Defending the UMBC network is a daunting challenge.  
There are over 10,000 users including over 500 faculty and staff members,
some of whose credentials are lost or compromised on a daily basis. 
Over 15,000 devices connect on any given day.
As a university, UMBC must support an open and collaborative 
environment---much more open than would be typically found at a commercial institution. 
In recent years, the main goal of attackers seems to be to compromise of personal data.
The Privacy Rights Clearinghouse estimates that, 
since 2010, educational institutions have experienced 411 data breaches 
impacting 18.7 million identities.\footnote{\url{https://www.privacyrights.org }}

For cybersecurity, DoIT employs five full-time professionals and 13 students, of whom four are graduate assistants. 
Their total annual budget for cybersecurity is approximately \$900,000.

\subsection{Origin and Use of the NetAdmin Tool} 
\label{sec:genesis}

DoIT created the NetAdmin tool circa 2005, primarily to automate the process of making certain exceptions to
firewall policies.  
Motivation for the NetAdmin tool was to provide the protection 
that firewalls and intrusion-prevention systems give commercial users,
while enabling researchers the ability to have an open network to perform their research 
unencumbered by network security devices. 
In 2006, when DoIT launched NetAdmin, UMBC was one of the first research universities to implement a default-deny firewall policy. 
Typically, less than 200 machines are exempted from firewall rules by faculty and staff.

Many researchers demanded the ability to connect from the outside, 
including via SSH, FTP, and HTTP, to their research machines 
which ran web servers and research tools.  
Before NetAdmin, researchers would submit a service ticket to be handled
on a case-by-case basis by DoIT staff, imposing significant delay and requiring DoIT staff hours.  Furthermore, 
this manual procedure made it difficult and error-prone to monitor and audit firewall exceptions, and to
terminate these exceptions upon expiration.

The real-time automatic NetAdmin tool addressed all of these issues, allowing faculty and staff 
to open the UMBC firewall 
to allow access from certain external systems to the machines they control on the research subnet.
The tool also greatly facilitated DoIT's ability to track firewall exceptions. DoIT used it extensively.

DoIT reasoned that the tool introduced little marginal risk to research machines
because machine owners could do whatever they pleased to their own machines.
As the students pointed out, however, there was risk to other UMBC systems, since the
possibility loomed that NetAdmin could introduce vulnerabilities allowing
an adversary to modify the UMBC firewall in a way that affected other machines, including 
machines outside of the research subnet.  

For over a decade, the tool ran untouched and seemed to work well.  There were no detected compromises, and state auditors
were satisfied with the process.  No one, however, had ever subjected NetAdmin to a critical technical security
evaluation. No attempt was ever made to update or patch any of the component subsystems on which NetAdmin depends.

Machines on the research subnet may have their own additional local firewalls, but most do not or do not have
them adequately configured.  
Such local firewalls might add some additional protection in depth.  Regardless, the issue of local firewalls
is mostly outside the scope of this study.

\subsection{Research Questions} 
\label{sec:questions}

During the four-day research study, students focused on the following questions.

\begin{enumerate}

\item What potential vulnerabilities (including from design, implementation, or configuration), if any, does NetAdmin introduce?

\item What potential attacks, if any, does NetAdmin enable?

\item What risks, if any, does NetAdmin introduce, especially 
beyond those present without NetAdmin?

\item Can an adversary---possibly a corrupt faculty or staff member---use NetAdmin to modify the firewall rules
for any machine outside of their control?

\item Is the NetAdmin architecture appropriate for its intended purpose?


\item Does NetAdmin use appropriate cryptographic functions
and does it use them appropriately?
Are the key lengths appropriate, and are keys managed properly?

\item Does NetAdmin use appropriate protocols, and does it
use them appropriately?

\item Is DoIT's policy involving NetAdmin clear and appropriate?

\end{enumerate}

Through discussions with DoIT in spring 2017, the study group came to a consensus on the choice of
the study topic.  DoIT suggested targeting NetAdmin and they suggested some of the initial
research questions, in much the way a client might engage a cybersecurity consultant for penetration testing.
Through further discussions and over the course of the study, students 
refined and augmented the initial questions.

In answering the research questions, the students could follow whatever approach they deemed fit, using whatever tools
they chose.  Throughout, the study group had access to DoIT staff who gladly provided any appropriate information requested.

The role of the study group was to advise DoIT on the facts concerning
potential vulnerabilities, attacks, risks, and mitigations.
As for the balance of risks and benefits of NetAdmin, that was a judgement only for DoIT to make
in consultation with the UMBC administration.
A compromise of computer systems or data could expose UMBC to liabilities, diminshed reputation, and recovery costs.

\subsection{Adversarial Model}
\label{sec:model}

We explain our adversarial model and trust assumptions.
In summary, our primary adversary is an outsider with compromised faculty or staff credentials
with the knowledge, skills, and resources of an excellent graduate student in computer science.

One way we characterize adversaries is by their level of access:
(1)~an outsider with compromised UMBC faculty or staff credentials,
(2)~a malicious faculty or staff member on the research subnet, and
(3)~a corrupt DoIT administrator.

The current situation offers no technical protection against a corrupt DoIT administrator:
DoIT manages the networks, the UMBC firewall, and NetAdmin, including the cryptographic keys and physical machines.
They control and monitor the log files.  Safeguards against corrupt system administrators 
depend mainly on observation, logs, and personal trust.
UMBC would, however, like to protect against Levels~(1) and~(2).

We also consider the adversary's increasing level of capability:
(a)~a ``script kiddie'' who can download and run basic malware and follow simple instructions,
(b)~a computer science graduate student with access to the university's computing resources, and
(c)~a nation state.
It would be impossible for UMBC to defend against a nation state, but 
UMBC would like to guard against Levels~(a) and~(b).\footnote{With the availability of increasingly 
sophisticated malware (e.g., flame), defending against even script kiddies is becoming a daunting challenge.}

We do not consider attacks on the underlying cryptography, the physical security of servers,
nor social engineering of DoIT staff.
Similarly, despite their practical importance, 
we do not analyze existing procedures nor their security implications
for recovery after disaster or compromise, including of the firewall API key or NetAdmin server.
We do, however, consider whether cryptography is properly used. 

The main goal of the adversary is to make unauthorized changes to
the UMBC firewall without detection.  The adversary may also wish
to use NetAdmin as a possible pivot for other attacks against the UMBC network.

\section{How NetAdmin Works} 
\label{sec:netadmin}

\begin{figure}[h]
\centering
\includegraphics[scale=1, bb=0 0 100 100]{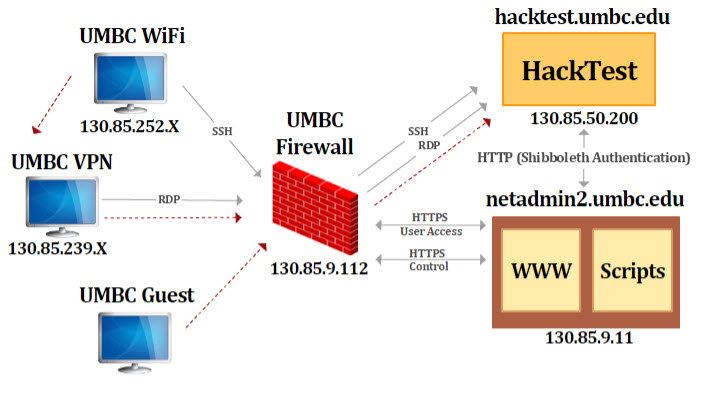}
\caption{Architecture of the ``HackTest'' virtual testing environment for the research subnet.}
\label{fig:hacktest}
\end{figure}

The NetAdmin tool is a custom web application that enables an authorized user to 
create and manage rules for the UMBC firewall that pertain to machines
on the UMBC research subnetwork under the researcher's control.
It is responsible for enforcing network policies on user-created firewall rules
and authenticating users during this process.
NetAdmin receives firewall rules from clients
and applies them to UMBC's Palo Alto firewall through API calls.\footnote{Application Programming Interface (API)}  
Additionally, UMBC network administrators frequently use {NetAdmin}
to maintain and audit firewall policies for a variety of campus network services. 
We explain how NetAdmin works by describing its
placement and implementation, authentication and authorization policies,
user interface, interaction with the UMBC firewall, and
method for storing firewall rules.

As shown in Figure~\ref{fig:arch}, NetAdmin resides on a dedicated
restricted server located on the UMBC network behind the UMBC firewall.
The server is configured to allow {NetAdmin} to be reachable 
only through connections from the campus network, 
including by clients originating from authenticated VPN connections. 
Clients wishing to make changes to the UMBC firewall directly in real time
communicate with the server through a web browser to create and modify firewall rules.
Alternatively, a client may send a request to DoIT; if the request is approved, a DoIT staff
member would implement the request using NetAdmin.
DoIT implemented {NetAdmin} in PHP 5.1.6, running on a 2.2.3 Apache HTTP server. 

NetAdmin recognizes several user groups including
full-time faculty, staff, and network administrator superusers.
These groups are defined in a file residing in the application directory on the NetAdmin server.
NetAdmin authenticates users and tokens they submit via UMBC's single sign-on Shibboleth service,
which is a Kerberos-like system.
Superusers may create and modify arbitrary firewall rules for any IP address on the UMBC network,
and they may view any firewall rule created by any user.
Faculty and staff users may create and modify some firewall rules for certain common ports
associated with their own network address.

Users specify firewall policies through a web-interface wherein the user's browser sends
web forms defining firewall rules to the NetAdmin server. 
A faculty or staff user may create, modify, or delete certain rules for any IP address they own. 
Rules that specify ports to open must reference ports on a list for commonly used services---for example,
SSH~(22), HTTP~(80), HTTPS~(443), and DNS~(53).
Proposed rules violating these restrictions must be submitted out-of-band to DoIT for special consideration.
After one year, rules expire and are marked ``inactive,'' regardless of use.
Users may activate, deactivate, or renew their rules through NetAdmin.

Through API calls invoked by Perl scripts, {NetAdmin} pushes rules to the UMBC firewall,
which is physically and logically separated from {NetAdmin}. 
The firewall authenticates API calls using a 360-bit symmetric API key 
stored in a file on the application directory of the NetAdmin server.  
The file is neither digitally signed nor hashed.

NetAdmin stores firewall rules and logs in unstructured files on the NetAdmin server.
The rules file helps preserve state through failures and restarts.
In this file, each rule is described by one record.
Newline characters delimit records, and pipe characters delimit fields within any record.  

Figure~\ref{fig:hacktest} shows the architecture of the virtual testing environment, called HackTest,
used in the study.  This environment includes guest UMBC accounts and allows access to all communications
between the UMBC firewall and the NetAdmin server.

\section{Potential Vulnerabilities, Attacks, and Risks}
\label{sec:results}

Our initial efforts focused on identifying potential vulnerabilities,
which we now summarize.  We also identify
potential attacks that exploit these vulnerabilities and 
discuss associated risks.

\subsection{Potential Vulnerabilities}
\label{sec:vulns}

\begin{enumerate}

\item The network architecture, and its use of a single firewall for the entire campus network, 
does not provide a segmented layer of defense for the research subnet.  \hbox{NetAdmin} implements a capability that 
is supposed to affect only the research subnet.  However, if an attacker could compromise NetAdmin, then
there is no architectural protection that would limit the attacker's ability to affect 
the entire campus network.

\item
The NetAdmin server runs on an unpatched, out-of-date operating system (OS)---CentOS~5.11---using Linux kernel (2.6.18),
which is known to have at least 451 vulnerabilities.\footnote{\url{https://www.cvedetails.com }} 
As of March 2017, CentOS~5.11 is no longer 
supported.\footnote{\url{https://wiki.centos.org/FAQ/General#head-fe8a0be91ee3e7dea812e8694491e1dde5b75e6d }} 
Compromise of the NetAdmin server could result in complete security failure 
of NetAdmin's firewall change process, allowing
an attacker to issue arbitrary firewall rules, modify log files, and exfiltrate the firewall API key.

\item
NetAdmin authenticates the firewall (using a self-signed certificate), but the firewall does not authenticate NetAdmin.  
The firewall checks only that forms purportedly from NetAdmin have the firewall API key.
Anyone with the firewall API key could issue arbitrary firewall rules.

\item
All communications between users and the NetAdmin server use unencrypted HTTP without integrity protection, allowing
an adversary to read and modify traffic and to carry out possible man-in-the-middle (MitM) attacks.
The user's browser sends NetAdmin rule changes via web forms, 
which the browser authenticates using UMBC's Shibboleth service.
By modifying the web forms sent from the user to the server, an adversary can
cause NetAdmin to implement firewall rules 
that enable unauthorized access to the user's machines.

\item
If an adversary could hijack an authorized user's entire session when the user logs into myUMBC, 
then they could masquerade as the user to NetAdmin.

\item 
While communications between the NetAdmin server and the firewall are protected by HTTPS,
there seems to be an opportunity for a possible additional MitM attack on the initial communication
that exchanges the self-signed authentication certificates 

\item
Because NetAdmin authenticates the firewall using a self-signed certificate, 
compromise of UMBC's key used to sign certificates would enable an adversary to
forge certificates.  

\item 
Committing some of the most common software security errors (see Kaza~\cite{kaza15}),
NetAdmin does not adequately check and sanitize inputs and form fields, allowing
the possibility of attacks that inject code and/or manipulate form field data.

\item
NetAdmin permits firewall rules to include HTML and JavaScript descriptions,
allowing possible code injection attacks~\cite{OWASP}.  For example, such injections might
permit the execution of arbitrary code in the browser of a user or system administrator
using NetAdmin.

\item 
NetAdmin does not check the length of firewall rule descriptions, allowing
possible record-overflow attacks and/or denial-of-service attacks (DoS).
In particular, NetAdmin's use of the PHP command fgetcsv() assumes that each record
is at most 999 bytes long.  The source code does not verify this assumption,
with the unintended result that any bytes following any 999-byte record will be
treated as a new record.

\item
Violating the principle of least privilege~\cite{bishop03}, 
NetAdmin uses a firewall API key with more permissions than are needed:
this key permits arbitrary changes to the firewall.

\item
NetAdmin stores critical information, including the firewall API key and log files, rules, 
in unstructured plaintext files on the NetAdmin server. 
The file is relatively easy to exfiltrate and is stored unencrypted.
The integrity of the file is not protected. 

\item
NetAdmin permits VPN access via the UMBC network,
facilitating remote attacks.

\end{enumerate}

\subsection{Potential Attacks}
\label{sec:attacks}

A variety of potential attacks are possible that exploit the identified vulnerabilities.  
We list a few examples.  Combinations of these attacks are also possible.  Because
NetAdmin allows VPN connections via the campus network, adversaries can mount many of these attacks remotely.
To demonstrate the feasibility of these attacks, students implemented 
a record-overflow and injection attack.

\begin{figure}[h]
\centering
\includegraphics[width=4in]{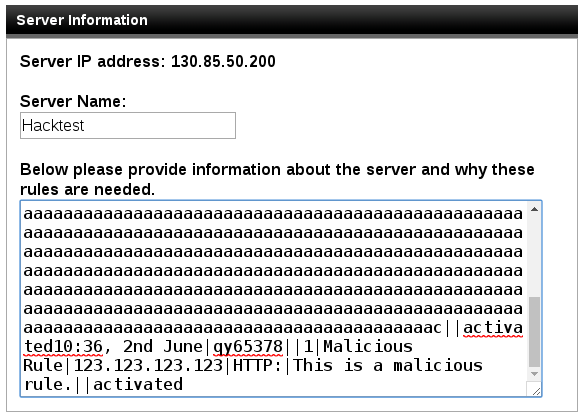}
\caption{Screenshot of NetAdmin web interface with record overflow.}
\label{fig:overflow}
\end{figure}

\begin{enumerate}

\item (record overflow)
An authorized user (or an adversary with compromised credentials of an authorized user)  
enters a long malicious rule that causes NetAdmin to implement an arbitrary rule for the 
UMBC firewall affecting any machine. Failure to check field lengths in user inputs results in NetAdmin
accepting data beginning with byte 1000 as a new and valid firewall rule.
See Figure~\ref{fig:overflow}.

\item (injection attack)
Similar to the record-overflow attack, but the payload is a firewall rule with malicious
HTML and/or JavaScript.  Victims can include users and administrators.  
The JavaScript can submit arbitrary rules to NetAdmin without the user knowing. 
It can submit rules via AJAX, and then when viewing the rules, 
simply delete that row from the HTML table, so the user never notices.
Alternatively, this attack might cause the web front-end to steal credentials 
or trick the victim to perform actions they do not wish to perform.
The malicious code can also execute arbitrary commands on the NetAdmin server, for example, 
exfiltrating the firewall API key. 

\item (server attack)
Exploiting known vulnerabilities of the Linux kernel version 2.6.18, the adversary learns the firewall API key and 
executes arbitrary commands on the NetAdmin server.  

\item (network attack)
Because the communications between users and NetAdmin are not protected for confidentiality or integrity,
a passive adversary can monitor all such communications.
An active adversary can modify messages to implement firewall rules affecting the user's machine
or to launch an injection or record-overflow attack described above.
For example, an adversary can mutate form fields to submit firewall rules that do not correspond to IP
addresses for which they are authorized.

\item (MitM session hijacking)
An adversary on the research subnet can attempt to hijack an authorized user's entire session as the user logs into
myUMBC.  For example, the adversary might try to do so by redirecting
the user's traffic to the adversary's machine using ARP poisoning or by posting a fake access point.
Easily available malware, such as 
SSL Strip, might faciliate such an 
attack.\footnote{\url{https://www.blackhat.com/presentations/bh-dc-09/Marlinspike/BlackHat-DC-09-Marlinspike-Defeating-SSL.pdf }}
Thereby, the adversary might be able to obtain a cookie that the adversary could use to authenticate to NetAdmin and to other
UMBC services.


\item (MitM attack during initial key establishment)
A sophisticated MitM attack between NetAdmin and the firewall can likely intercept
the the firewall API key and cause NetAdmin to accept an erroneous public-key certificate of
the firewall.  This attack requires access at the crucial key-establishment time. 

\end{enumerate}

\subsection{Potential Risks}
\label{sec:risks}

The attacks listed above are feasible and well within the capabilities of computer science graduate students.
The adversary could learn the firewall API key, from which the adversary could implement arbitrary rules on the
UMBC firewall.  Thus, the adversary could open any port for any machine on the UMBC firewall.
By opening ports, the adversary might move throughout the campus network, 
exfiltrate data, steal credentials, steal intellectual property, sabotage research, disrupt services, 
and launch DOS attacks.
By manipulating log files on NetAdmin, the adversary could try to remain unnoticed.

The UMBC network is under constant attack. 
DoIT believes that most of the current attacks aim to steal personal information, including credit card numbers.
DoIT is not aware of any attack having been mounted on NetAdmin.  
Perhaps this situation reflects that no adversary bothered to explore possible vulnerabilities 
in NetAdmin and that there may be other more attractive avenues of attack.

\section{Recommendations}
\label{sec:mitsandrecs}

To address the potential vulnerabilities, attacks, and risks
identified in Section~\ref{sec:results}, the students 
recommend that DoIT carry out the following mitigations 
and additional actions.

\subsection{Recommended Mitigations}
\label{sec:mits}

\begin{enumerate}

\item (keep software up-to-date)
The NetAdmin server should run a well-patched and maintained operating system.
For example, replace CentOS 5.11 (which is deprecated)
with the most recent version (CentOS~7 as of March 2018). 
All software used by NetAdmin, particularly the Apache HTTP server, PHP engine, and Perl interpreter,
should be up-to-date. 
Doing so reduces the ability of an adversary from using easily available known attacks.

\item (sanitize inputs)
Sanitize all form inputs on the server side. 
Special characters, particularly the pipe character used to delimit records in the data file, should
be prohibited in firewall rule descriptions. Form fields should not accept HTML or Javascript
values.

\item (check IP addresses)
For non-superusers, the server should
ensure that the IP address of a firewall rule corresponds to the IP address of the user. 
Although an adversary might be able to spoof her IP address, it would be prudent to
add this check.

\item (check record sizes)
In firewall rules, enforce maximum sizes for fields including server names and rule descriptions.
For example, one might impose a 128-byte limit on server names and a 256-byte limit on rule descriptions.
Doing so prevents an adversary from mounting record-overflow attacks to create arbitrary firewall rules.
Read data files in such a way that records larger than 999 bytes are rejected, or truncated, 
rather than interpreted as multiple records.

\item (encrypt communications)
To protect confidentiality and integrity, 
communications between the NetAdmin server and NetAdmin users should use end-to-end encryption
with authentication and integrity protection, 
such as that offered by implementations of OpenSSL.

\item (limit capabilities of firewall API key)
Limit the capabilities of the API key used to communicate with the UMBC firewall
so that it can only create, modify, and delete
firewall rules pertinent to the UMBC research subnet. 
Different API keys should be used for
pushing firewall rules defined by users versus pushing firewall rules defined by superusers. 

\item (improve key management)
The Firewall and NetAdmin each should have a certificate signed by a trusted third party.
Each machine should authenticate the other.
Improve the initial establishment and storage of the firewall API key by NetAdmin.
Use a key-establishment protocol that is not vulnerable to MitM attack.
Do not store the API key in plaintext in an unprotected file.  
Consider using a TPM.\footnote{Trusted Platform Module (TPM)}
Some database products might provide some degree of protection for key storage.

\item (deny VPN access to NetAdmin)
Require all connections to NetAdmin to originate from campus, as enforced in part
through physical network connections to the NetAdmin server.
To reduce the risk of remote attacks, VPN access to NetAdmin should be denied.
Doing so, however, would still allow the possibility of a user connecting
to NetAdmin indirectly after remotely connecting to another campus machine,
an unfortunate reality that would be hard to prevent.

\item (segment NetAdmin)
For defense-in-depth, segment NetAdmin into two services: a web front-end, and a back-end
that communicates with the firewall.  The front-end accepts input, validates whom it came from,
sanitizes it, and passes it to the back-end.  The back-end performs additional validation and 
sends instructions to the firewall.  Only the second service is permitted to read sensitive 
data (e.g., firewall API key).  The services should run under separate accounts.
The front-end should be unable to run shell scripts, and the back-end should be permitted
to communicate only with the front-end and the firewall.

\end{enumerate}

\subsection{Additional Recommendations}
\label{sec:recs}

\begin{figure}
\centering
\includegraphics[scale=1, bb=0 0 100 100]{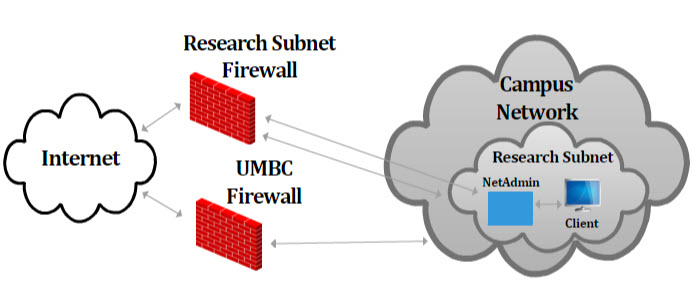}
\caption{Recommended architecture to provide compartmentalized defense, architecturally limiting 
the scope of NetAdmin functionality to the research subnet.}
\label{fig:rec-arch}
\end{figure}

\begin{enumerate}

\item (deploy a segmented architecture)
To limit the possible scope of failures in NetAdmin, deploy a segmented architecture as shown
in Figure~\ref{fig:rec-arch}, so that the architecture will prevent NetAdmin from affecting
any machine outside of the research subnet.  

\item (physically secure NetAdmin and firewall in same room) 
Improve the security of the connection between NetAdmin and the proposed separate firewall
by placing both machines in the same physically-secure room physically connected by a dedicated subnet.
Enforce this architecture in part through
physical network connections between the NetAdmin server and the proposed 
separate firewall for the research subnet.

\item (conduct technical security audits)
In addition to periodic administrative audits, DoIT should also commission technical security audits
carried out by qualified independent cybersecurity consultants.

\end{enumerate}

\section{Discussion}
\label{sec: discuss}

We now discuss a number of issues raised by our experiences, including
vulnerabilities discovered, 
takeaways for study organizers, 
UMBC's extension of SFS scholarships to two community colleges, and
reflections from a UMBC system administrator.

\subsection{Observations on Vulnerabilities Discovered}

Our case study exposes why certain common
vulnerabilities arose and the costs and complexities needed to
avoid or mitigate these vulnerabilities.  
For example, the NetAdmin server ran an obsolete and unpatched OS
partly because DoIT did not invest the staff resources to
keep the system current.  DoIT feared that updates to the OS might
break the custom system, requiring more staff time to fix.  
There is no escaping the harsh reality that maintaining a secure
system demands considerable ongoing attention and resources, 
and any changes to the system (including ones mitigating known vulnerabilities) 
also risk the possibly of introducing new vulnerabilities. 

\subsection{Takeaways for Study Organizers}

Overall the study went very smoothly.  
Project-based learning sustained inquiry and critical thinking.
The virtual environment enabled the students to work on an essentially identical copy
of the network environment without interfering with the actual network.
The evening chat sessions provided a way for students to participate, even if they
could not attend the regular session.  We found that chat worked better than
video conferencing because it created a written record and it more easily enabled
students to enter the session late.
Using the study room's data projector and whiteboard facilitated discussions.
Throughout each day, the group posed questions to DoIT and received written answers
via a GoogleDoc.
Talking with the developer was very useful.
The study room could have benefited from more electrical outlets and power strips.

Participants completed surveys at the end of the summer project in which 
they all reported that the project increased their cybersecurity knowledge and skills 
(86\% responded strongly agree, 14\% responded agree).   
Participants identified the following elements as valuable:  
teamwork, hands-on nature of the task, real-world challenge, critical thinking, and problem solving.  
All participants reported that they would recommend the summer study project to other cybersecurity students.

At UMBC we were fortunate to enjoy strong support and cooperation from DoIT.  
Some other schools, however, might face a defensive administration fearing embarrassment 
or unwilling to trust students with sensitive system security information.
We commend UMBC's DoIT for their constructive attitude, which through encouraging 
analysis of their systems, helps them enhance the security of their operations.
Most SFS students are highly trustworthy:
schools select SFS students carefully, and the students are expected to be able to attain at least
a secret clearance (many hold top-secret clearances).  

Our choice to schedule the study in early summer immediately after
commencement created a conflict with several students who were
starting required summer internships.  To avoid such conflicts, this
year we held the study in January, the week before classes resume.
While better for most students, this new schedule created conflicts with
some students who took classes in January.

\subsection{Extending SFS Scholarships to Community Colleges}

So far, the new venture to extend SFS scholarships to community colleges appears to be
working well.  We are attracting highly-qualified students, and for some students from modest backgrounds, 
the scholarship is a life-changing opportunity.  There is an opportunity cost in that, given finite budgets,
a scholarship awarded to a CC student who transfers to UMBC is a scholarship not awarded to a non-transfer student at UMBC.
Awarding one SFS scholarship to each of two CCs per year seems like an appropriate balance for UMBC.

Among CC students, we restrict our attention to CC students pursing Associate's (AS) degrees who intend to transfer to a four-year school.
UMBC's articulation agreement with Maryland CCs ensures that any student who earns an AS degree at
any Maryland CC with GPA of at least 3.0 will be admitted to UMBC.
There remains the issue, however, that many CCs target their cybersecurity courses at 
Associate of Applied Science (AAS) students who
intend to enter the workforce after two years at CC.  
Typically, the AAS students are not prepared
to transfer to four-year schools, and often AS students have limited available time to take many
cybersecurity courses.

\subsection{Reflections from a UMBC System Administrator}


Jack Suess,
Vice President of Information Technology and Chief Information Officer (CIO) at UMBC, remarks:

\begin{quote}
As the CIO and a former student IT employee (1979--81), I greatly value this experience for our staff and students. 
For staff, we are forced to move out of our imagined protective bubble and 
look at the consequences of our decisions if we have a motivated and highly skilled attacker. 
In this case we can ``imagine'' that no one will ever take the effort to dig deep enough to find our flaws; 
when the students unmask these flaws, it shows that we need to be more conscious of flaws! 
For students, they are able to leave the academic bubble of the theoretical and operate 
in the real world of people writing code that needs to be secure and to think about what can go wrong. 
This effort will make those who program much better at thinking about security from the beginning of their design. 
As a CIO, I love that we have repeated this exercise with the SFS scholars 
and, in both exercises, we gained valuable feedback that improved the security of our services for free!
\end{quote}

\section{Conclusion}
\label{sec:conclude}

In only four days, SFS scholars in the UMBC summer study identified significant security
vulnerabilities with UMBC's NetAdmin system 
that enables faculty and staff to open the UMBC firewall
for machines they control on the research subnet.
These vulnerabilities include unpatched systems, record overflow, 
uncompartmentalized architecture, and failure to sanitize inputs.
Students also suggested mitigations. 
The details provide an instructive authentic case study in cybersecurity.

As evidenced by responses from the student exit questionnaire, the experience
engaged and motivated students. The participants who had transitioned from community
colleges demonstrated that there are highly capable students at community colleges
who can contribute to our nation through cybersecurity.

The study illustrated advantages that can arise when educators partner with
university system administrators 
to allow qualified students to use the university's computer systems as a learning laboratory.   
Students gained exciting, concrete, hands-on collaborative experience.
Educators were given a rich and inspiring detailed realistic case study to support project-based learning.
The university's system administrators 
received a free cybersecurity consultation and 
hired several of the SFS scholars to join their security team.

We look forward to continuing the research study experience each year, and 
we hope that other schools can also benefit from such collaborations.

\section*{Acknowledgments} 

We thank Jack Suess and Damian Doyle (UMBC Division of Information Technology)
for their enthusiastic cooperation and for providing a virtual machine of
the software environment.   
We thank Michael Oehler and Edward Zieglar (National Security Agency)
for interacting with the study participants.
Mykah Rather drew the figures.
Peter Peterson and Travis Scheponik offered helpful comments.
Thanks also to programmer Ray Soellner (UMBC) for bravely 
attending daily briefings.
This project was supported in part by the
National Science Foundation under SFS grant 1241576.
Sherman was also supported by the
U.S. Department of Defense under 
CAE-R grant H98230-17-1-0349 and 
IASP grant H98230-17-1-0387.

\bibliography{sfs-ss-bib}
\bibliographystyle{alpha}

\vfill
\appendix
\section{Acronyms and Abbreviations}

\begin{tabular}{ll}
API & Application Programming Interface\\
AS & Associate of Science\\
AAS & Associate of Applied Science\\
CAE & National Center of Academic Excellence\\
CC & Community College\\
CCDC & National Collegiate Cyberdefense Competition\\
DoD & Department of Defense\\
DoIT & UMBC Department of Information Technology\\
GPA & Grade Point Average\\
IASP & DoD Information Assurance Scholarship Program\\
INSuRE & Information Security Research and Education collaborative\\
IT & Information Technology\\
MC & Montgomery College\\
MitM & Man-in-the-Middle\\
NJIT Cyber-RWC & New Jersey Institute of Technology Cybersecurity Real World Connections\\
NSA & National Security Agency\\
NSF & National Science Foundation\\
OS & Operating System\\
PBL & Project-Based Learning\\
PGCC & Prince George's Community College\\
SCAMP & Summer Conference on Applied Mathematical Problems\\
SFS & CyberCorps: Scholarship for Service\\
TPM & Trusted Platform Module\\
UMBC & University of Maryland, Baltimore County\\
VPN & Virtual Private Network
\end{tabular}

\bigskip \bigskip \bigskip \noindent
{Full report submitted to {\it Cryptologia} (\today).}


\end{document}